\documentstyle [12pt]{article}
\begin{document}
\newcommand{\vm}{\vspace{0.2cm}}
\newcommand{\vl}{\vspace{0.4cm}}

\title{Diff
invariant Poincare transformations as deformation of Poincare algebra. }
\author{Matti Pitk\"anen\\
Torkkelinkatu 21 B 39, 00530, Helsinki, FINLAND}
\date{27.10. 1994}
\maketitle

\newpage

\begin{center}
Abstract
\end{center}

\vm

The concept of $Diff^4$ invariant Poincare transformations is a cornerstone
of  T(opological) G(eometro)D(ynamics). This concept makes it possible to
understand the concept of subjective time and irreversibelity as well as
nontriviality of S-matrix at  quantum level. In this paper the possibility
of  identifying  $Diff^4$ invariant Poincare transformations as  the
recently discovered  Lorentz invariant deformation of Poincare algebra with
the basic property that 'new' energy is some function of 'old' energy, is
considered.

\newpage

  \section{Diff
invariant Poincare transformations as deformation of Poincare algebra.}

One of the cornerstones of TGD is the concept of $Diff^4$ invariant Poincare
 transformations described in \cite{TGD}: the appendices of $\cite{TGD}$
contain additional developments related to the concept. The concept arises
in the following manner. \\ a) The configuration space of the  TGD consists
of all 3-surfaces in the cartesian product of future light cone $M^4_+$
(points $m^k$ of $M^4$ satisfying  $m_{kl}m^km^l=(m^0)^2- (\bar{m})^2\ge 0,
m^0\ge0$).  \\ b) K\"ahler geometry for this space is defined  in terms of
K\"ahler function, which corresponds to the absolute minimum for so called
K\"ahler action, which is $Diff^4$ invariant.   This definition associates to
each 3-surface  $X^3$ a unique spacetime surface $X^4(X^3)$,  the classical
history associated with 3-surface.  In \cite{padTGD} it is explained how
this concept defines quantum counterpart of the Thom's catastrophe theory:
in this theory discontinuous jumps take place along 'Maxwell line' rather
than along the 'fold line':  this is  what is known to happen  in
phase transitions.  \\ c) K\"ahler action is same for 4-surface and its
Poincare translate but Poincare transformations to not  in general map
absolute minimum to absolute minimum: exception is formed by the Lorentz
transformations mapping light cone to itself.  Therefore Poincare invariance
is broken and ordinary representations  of Poincare group are not $Diff^4$
invariant.  \\ d) One can overcome the problem by modifying the concept of
Poincare invariance.  By $Diff^4$ invariance of state functions one can
consider instead of  3-surface $X^3$   the 3-surface $X^3_a$, the
intersection $X^4(X^3)$ with  the cartesian product of  light cone proper
time constant hyperboloid  $H_a=\{m^k\vert  m_{kl}m^km^l=a^2\}$ with
$CP_2$.  $X^3_a$ is invariant under Lorentz group and one can define the
action of infinitesimal Poincare transformation by requiring that the
action on $X^3_a$ is  ordinary  infinitesimal Poincare transformation:  the
action on other 3-surfaces on $X^4(X^3)$ is fixed by the requirement that
$X^4(X^3)$ is replaced with absolute minimum surface associated with the
infinitesimal Poincare translate of $X^3_a$. The result is that absolute
minimum surface suffers nontrivial deformation and even its topology can
change.  \\ e) This $Diff$ invariant realization of Poincare invariance
depends on the value of proper time parameter $a$ and continuous family of
nonidentical  unitarily related energy momentum eigenstate basis  are
obtained.  A natural interpretation for proper time parameter $a$ is as
subjective time experienced or measured by conscious observer.  The arrow of
time has simple geometric  explanation: there is much more room in the
future than in past for a point inside future light cone and quantum
transitions tend to increase the value of the  proper time variable in the
long run.

\vm

There are however open questions. What happens to the Poincare group?
Commutation relations get deformed but do they close anymore?  If not should
one try to extend the Poincare algebra in order to get a closed algebra?
Could quantum groups have some relevance in the problem? The answers to
these questions seem to be beyond calculational capacities  since it is
difficult to imagine how  one could  deduce analytic expressions for the
action of $Diff^4$ invariant  Poincare transformations for such a
complicated structure as the space of all possible 3-surfaces in
$M^4_+\times CP_2$   is.

\vm

A great surprise  in this respect was the paper of Kehagias and Meessen,
where it was shown that Poincare group allows deformation with exact Lorentz
algebra:  the structure in accordance with  the most optimistic expectations
and raises the  concept of $Diff^4$ invariant Poincare transformations  on
sound footing!  What happens in deformation is a modification for the
expression of energy $P_0(diff)$:  the new energy is certain function $\beta
(P_0)$ of the  'old' energy $P_0$!   The old energy corresponds to the energy
associated with ordinary Poincare transformations and new energy to the
energy associated  with $Diff^4$ invariant Poincare transformations.
Lorentz invariance corresponds to the Lorentz transformations leaving the
future light cone invariant. The only thing, which one cannot calculate at
this stage is the explicit form of the function.

\vm

It is worthwhile to write the expressions for the deformed commutation
 relations

\begin{eqnarray}
Comm(J_i,J_j) &=& i\epsilon_{ijk}J_k\nonumber\\
Comm(J_i,K_j) &=& i\epsilon_{ijk}K_k\nonumber\\
Comm(J_i,P_j) &=& i\epsilon_{ijk}P_k\nonumber\\
Comm(K_i,K_j) &=& i\epsilon_{ijk}J_k\nonumber\\
Comm(J_i,P_0) &=& 0\nonumber\\
Comm(P_i,P_j) &=& 0\nonumber\\
Comm(K_i,P_0) &=&  i\alpha (P_0) P_i\nonumber\\
Comm(K_i,P_j) &=& i\beta (P_0)\delta_{ij} \nonumber\\
\alpha (P_0)\frac{d\beta (P_0)}  {dP_0}&=&1
\end{eqnarray}

\noindent Deformation  is trivial at the limit

\begin{eqnarray}
\alpha (P_0)&=&1 \nonumber\\
\beta (P_0)&=&P_0
\end{eqnarray}

\noindent  The deformed algebra leaves invariant the lenghts of deformed
four momentum vector and Pauli-Lubanski vector

\begin{eqnarray}
m^2&=& m_{kl}P^k(d)P^l(d) \nonumber\\
W^2(d)&=& m_{kl}W^k (d)W^l(d)\nonumber\\
P_k (d) &\leftrightarrow & (\beta (P_0),P_k)\nonumber\\
W_i (d) &\leftrightarrow& (W_0=\bar{J}\cdot P, W_i= \beta (P_0)J_i+
 \epsilon_{ijk}P_jK_k  )\nonumber\\
\
\end{eqnarray}

\noindent    The effect of deformation is clearly to replace the expression
 of energy with more general one.

\vm

Consider first  the  possible  application  of the concept to Quantum TGD at
 particle physics length scales. \\
a) The nontriviliaty of S-matrix in Quantum TGD follows from the parametric
 dependence of $\beta (P_0)$ on  light cone proper time $a$. There are good
reasons to expect that this dependence is extemely weak at Planck length
scales. Although the
deviation from triviality might be extremely small the criticality of TGD
Universe at quantum level is expected to imply initial value sensitivity and
large deviations   of S-matrix  from unit matrix in particle physics length
scales. The essential point is the unstability of 3-surface (particle) to
topological decay  to several 3-surfaces: only a small deformation (say
small time translation) can cause this decay.     \\  b)  Particle physics
applications \cite{padTGD,padmasses} necessitate the consideration of p-adic
field theory limit and an there are no obstacles of generalizing the
deformation to p-adic case.  The correspondence between p-adic and real
numbers is given by $\sum_nx_np^n \rightarrow \sum_nx_mp^{-n}$ so that $x$
and $x+p^n$ are numbers near to each other for large and positive value of
$n$.    In particle physics applications $\beta (P_0)$ should not deviate
much from $P_0$ and a natural expansion parameter is the p-adic prime $p$ so
that one can write

\begin{eqnarray}
\beta (P_0)&=& P_0 (1 + p f(P_0))\nonumber\\
f(P_0)&=& f_0 +O(p)
\end{eqnarray}

\noindent      If $f_0$ is integer  corrections to ordinary dispersion
 relation are extremely small for primes related to the elementary particle
mass scales, typically Mersenne primes $M_n=2^n-1$ with $n=127,107,89$
\cite{padmasses}.   If $f_0$ is rational number corrections to energy are
typically below  one per cent \cite{padmasses}.

\vm

In long length scales (macroscopic, astrophysical, cosmological) the
difference between Poincare energy and $Diff^4$ invariant Poincare energy
could be large.
  \\ a)
$Diff^4$ invariant energy vanishes,  when absolute minimum 4-surface is
static since time translations correspond to  $Diff^4$ transformations and
must leave state invariant.    For small periodic deformations Diff
invariant energy is just a multiple for the frequency of oscillation by same
argument and one obtains just harmonic oscillator. For nonperiodic surfaces
near these simple surfaces the previous equation makes possible to $Diff^4$
invariant energy.    These observations  suggest that for simple systems
$Diff^4$ invariant energy is roughly the  vacuum subtracted energy, vacuum
energy defined as the ordinary Poincare invariant energy of static
configuration. \\ b)   The function $\beta (P_0)$,  the deformed energy
can    depend on 3-surface via Poincare invariant parameters describing
general  geometric properties of the 3-surface such as form and size.   The
considerations of appendix of \cite{padTGD} suggest that  $\beta$ depends on
3-surface.   For  example, for TGD:eish cosmic strings \cite{TGD} having
string tension of order $1/G$,   Poincare invariant energy vanishes in
static case  unlike the ordinary Poincare energy, which  for string section
of length  $L$ is of same order of magnitude as mass of black hole with
Schwartshild radius $L$!  For  periodically rotating cosmic strings $Diff^4$
invariant energy  is essentially the kinetic energy of rotational motion
with rest mass excluded and this can explain the reduction of  effective
string tension by a factor of order $10^{-5}$ from the value $T\sim 1/G$
for  cosmic strings used to model galaxy formation and to explain dark
matter puzzle \cite{TGD}. \\
c)  At the classical limit of the theory  it should be generally
possible to identify the eigen value of quantum energy $\beta (P_0)$ with
its classical counterpart  $\beta (P_0)$ , where $P_0(X^3)$ is  the
classical conserved Poincare
energy associated with the absolute minimum surface $X^4(X^3)$.   For state
functionals dispersed in a set of 3-surfaces with different values of
$P_0(X^3)$ the identity of classical and quantum energies requires that  the
functional form of the classical counterpart for the Poincare invariant
energy  $\beta (P_0 (X^3))$  must depend  on 3-surface in
the manner dictated by the condition

\begin{eqnarray}
\beta (P_0 (X^3),X^3)&=& \beta_0\equiv \beta (P_0 (X^3_0),X^3_0)
\end{eqnarray}

\noindent  The parametric dependence is only on those parameters, which are
 Poincare invariant and the two dependences on $X^3$  must compensate  each
other in the expression of energy.  One must  fix the value of $\beta_0$ by
some criterion for some surface.  A reasonable choice is surface, which
corresponds  periodic oscillation around static
ground state, for which $Diff^4$ invariance implies gives
 oscillator spectrum. \\  d) An interesting  (but perhaps purely
formal) possibility is that the dispersion relation $P_0=
m+\frac{\bar{p}^2}{2m}$  characteristic to Galilei invariance and resulting
approximately at nonrelativistic limit is in fact exact relationship implied
by suitable deformation of Poincare algebra

\begin{eqnarray}
\beta^2 (P_0) &=& 2mP_0- m^2 \ge 0   \nonumber\\
\alpha (P_0)&=& \frac{\sqrt{\beta}}{m}
\end{eqnarray}

\noindent  Effective  Galilei invariance would result from the deformation
 of Poincare group.   This deformation depends on particle mass but this is
in accordance with the dependence of $\beta$ on 3-surface since   each
particle corresponds to certain classical surface at semiclassical limit of
TGD.  The  reduction of ordinary Poincare invariance to effective Galilei
invariance (but «cosmological« Lorentz transformations acting as exact
symmetries!)  would be  implied basically by the  absolute minimum
requirement for K\"ahler action.

\end{document}